\documentclass[journal,twoside,web]{ieeecolor}
\usepackage{generic}
\usepackage[utf8]{inputenc}
\usepackage{cite}
\usepackage{amsmath,graphicx}
\usepackage{mathtools}
\usepackage{multirow}
\usepackage{subfigure}
\usepackage{setspace}
\usepackage{hyperref}
\usepackage{color,soul} 
\usepackage{tabularx} 
\usepackage[ruled,vlined]{algorithm2e}

\newcolumntype{R}[1]{>{\raggedleft\arraybackslash}p{#1}} 
\DeclareMathOperator*{\argmin}{arg\,min}

\newcommand*{\defeq}{\mathrel{\vcenter{\baselineskip0.5ex \lineskiplimit0pt
                     \hbox{\scriptsize.}\hbox{\scriptsize.}}}%
                     =}

\title{App-based saccade latency and error determination across the adult age spectrum}
\author{Hsin-Yu Lai, \IEEEmembership{Student Member, IEEE}, Gladynel Saavedra-Pe\~na,  Charles G. Sodini, \IEEEmembership{Fellow, IEEE}, \\Thomas Heldt, \IEEEmembership{Senior Member, IEEE}, Vivienne Sze, \IEEEmembership{Senior Member, IEEE}
\thanks{This work was supported in part by Sensetime through a grant to the MIT Quest for Intelligence and by the MIT-IBM Watson AI Lab.}
\thanks{H.-Y. Lai is with the Department of Electrical Engineering \& Computer Science, Massachusetts Institute of Technology, Cambridge, MA 02139, USA.} 
\thanks{G. Saavedra-Pe\~na was with the Department of Electrical Engineering \& Computer Science, Massachusetts Institute of Technology, Cambridge, MA 02139, USA.}
\thanks{C.G. Sodini is with the Department of Electrical Engineering \& Computer Science, Institute for Medical Engineering \& Science, and the Microsystems Technology Laboratory, Massachusetts Institute of Technology, Cambridge, MA 02139, USA.}
\thanks{T. Heldt is with the Department of Electrical Engineering \& Computer Science, Institute for Medical Engineering \& Science, and the Research Laboratory of Electronics, Massachusetts Institute of Technology, Cambridge, MA 02139, USA.}
\thanks{V. Sze is with the Department of Electrical Engineering \& Computer Science and the Research Laboratory of Electronics, Massachusetts Institute of Technology, Cambridge, MA 02139, USA. e-mail: sze@mit.edu}
}
\date{\today}
\begin{document}

\maketitle
\begin{abstract}
\textit{Objective}: We aid in neurocognitive monitoring outside the hospital environment by enabling app-based measurements of visual reaction time (saccade latency) and error rate in a cohort of subjects spanning the adult age spectrum.
\textit{Methods}: We developed an iOS app to record subjects with the frontal camera during pro- and anti-saccade tasks. We further developed automated algorithms for measuring saccade latency and error rate that take into account the possibility that it might not always be possible to determine the eye movement from app-based recordings.
\textit{Results}: To measure saccade latency on a tablet, we ensured that the absolute timing error between on-screen task presentation and the camera recording is within 5 ms. We collected over 235,000 eye movements in 80 subjects ranging in age from 20 to 92 years, with 96\% of recorded eye movements either declared good or directional errors. Our error detection code achieved a sensitivity of 0.97 and a specificity of 0.97. Confirming prior reports, we observed a positive correlation between saccade latency and age while the relationship between error rate and age was not significant. Finally, we observed significant intra- and inter-subject variations in saccade latency and error rate distributions, which highlights the importance of individualized tracking of these visual digital biomarkers.
\textit{Conclusion and Significance}: Our system and algorithms allow ubiquitous tracking of saccade latency and error rate, which opens up the possibility of quantifying patient state on a finer timescale in a broader population than previously possible.

\end{abstract}

\vspace{-10pt}
\section{Introduction} \label{sec:introduction}

It remains challenging to track neurodegenerative disease progression objectively, accurately, and frequently. Current assessments of neurodegenerative diseases are subjective and sparse, and standard neurocognitive and neuropsychological test batteries require a trained specialist to administer and score~\cite{Hoops_2009,Mitchell_2009}. Additionally, these tests demand significant patient time and cooperation, and can therefore be influenced by a patient's level of attention and comfort with the clinical setting~\cite{NAP_2018}.
Quantitative, objective, and frequent assessments may mitigate the effects of individual physician’s clinical acumen and patient fatigue when determining neurodegenerative disease progression. 

Assessment of eye movement is a promising candidate for such a quantitative and objective test.  First, eye movements are readily observable. Second, their neural pathways involve several brain regions, and they might hence be affected by degenerative processes affecting various brain centers~\cite{Leigh_2015}. 
For example, Huntington's disease and progressive supranuclear palsy directly affect oculormotor pathways. As a result, clinical eye movement assessments are key to diagnosing and tracking these diseases.

In the context of neurodegenerative disease assessment and progression monitoring, pro- and anti-saccade visual reaction tasks are often used challenge tests~\cite{Tabrizi_2009,Anderson_2013}. In the pro-/anti-saccade tests, a subject is asked to look towards/away from a visual stimulus. 
An anti-saccade task, in particular, requires a person to inhibit a natural reflexive eye movement towards the stimulus and initiate an eye movement in the opposite direction of the stimulus. Thus, it requires more cognitive processing than a pro-saccade task~\cite{Munoz_2004,McDowell_2008}. 
Because these tasks demand cognitive abilities which can be affected by neurodegenerative diseases, two saccadic eye movement features were observed to be significantly different between healthy subjects and patients: saccade latency (visual reaction time) and error rate (the proportion of eye movements towards the wrong direction)~\cite{Shafiq_2003,Crawford_2005,Mosiman_2005,Garbutt_2008}. However, these features are commonly measured with dedicated infrared cameras and chinrests, which limits the measurements to the doctor's office or the neurophysiological laboratory.  
In our previous work \cite{Lai_2018,Saavedra_2018,lai_2020}, we showed that we can accurately and robustly determine saccade latency from recordings obtained with a smartphone camera. 

In this work, we developed an app to display the pro-/anti-saccade tasks on a tablet computer while recording a subject's eye movements with the built-in camera. 
We present an automated processing pipeline to determine pro-/anti-saccade latency and error rate, thus enabling ubiquitous recording of these neurological digital biomarkers. With this novel recording platform and pipeline we collected over 6,800 videos and over 235,000 individual eye movements from 80 subjects across the adult age spectrum.
\section{Materials} \label{sec:materials}

\subsection{Recruitment}\label{sec:recruitment}
To study the responses of subjects of different ages to pro- and anti-saccade tests, we recruited 80 self-reported healthy adult subjects, ranging in age from 20 to 92 years.
\begin{figure}[b]
      \centering
      \includegraphics[width=\columnwidth]{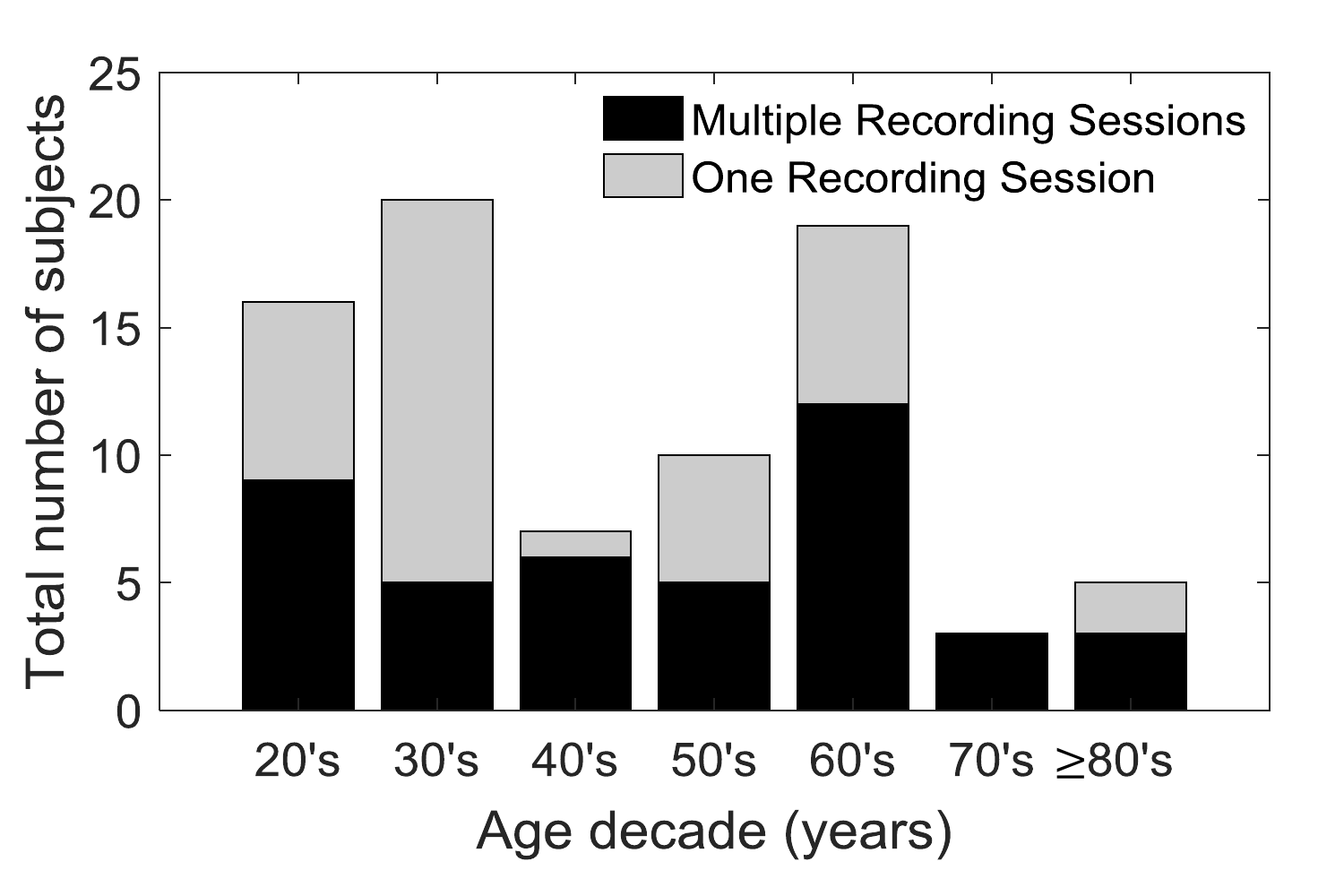}
     
      \caption{{Age distribution of subjects with single or multiple recording sessions.} 
      }
       
      \label{fig:subject}
  \end{figure}

Video recording of volunteers was approved by MIT’s Committee on the Use of Humans as Experimental Subjects (protocol \# 1711147147), and informed consent was obtained from each participant before recording. 
A subject could choose to participate either once or for multiple recording sessions.
Subjects who chose to participate in a single session recorded three pro-saccade tasks and three anti-saccade tasks, where each task consists of a set of 40 stimuli. Subjects who chose to participate in multiple recording sessions were asked to take three pro-saccade and three anti-saccade tasks every day for at least two weeks and were given the choice of 20 or 40 stimuli per task. 
The number of single and multiple recording sessions, grouped by decades of age is shown in Fig.~\ref{fig:subject}.

\subsection{App design} \label{sec:app-design}
In our previous work~\cite{lai_2020}, we displayed the visual reaction task on a laptop and recorded the subjects with an iPhone. Synchronization of the recording and task display was achieved through a second screen that mirrored the laptop screen and was recorded alongside the subject's response~\cite{lai_2020}. Given the elaborate set-up, the recording was limited to our laboratory setting.
Our goal here was to allow for ubiquitous recording and hence for subjects to record themselves in the comfort of their homes or offices. 
We therefore developed an iOS app so subjects could record themselves with the frontal (i.e. selfie) camera as the tasks were displayed on the screen. While the app can run on iPhones, our platform of choice was the iPad (Generation 2 and 3) for their larger dimensions and hence larger angular gaze amplitudes ($\sim$12.7 degrees at a distance of 40 cm to the camera).

\begin{figure}
      \centering
      \includegraphics[width=\columnwidth]{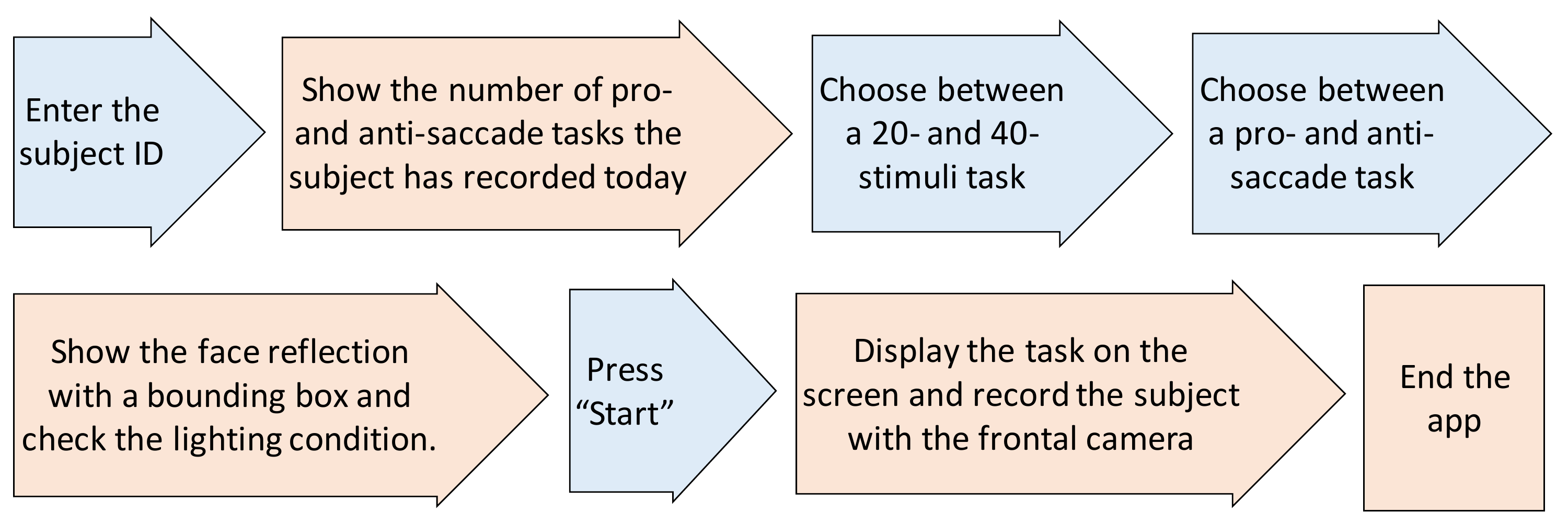}
     
      \caption{{The flow of the app. Blue arrows request the input from the subject. Orange arrows denotes the response of the app.} 
      }
       
      \label{fig:app-chart}
  \end{figure}

The flow of the app is shown in Fig.~\ref{fig:app-chart}. The app first obtains the subject's ID and then reminds the subject of the number of pro- and anti-saccade tasks they have performed the same day. Subsequently, subjects are prompted to select the number of stimuli they wish to perform (20 or 40) and whether they would like to perform a pro- or anti-saccade task.

\begin{figure}
      \centering
      \vspace{-9pt}
      \includegraphics[width=0.8\columnwidth]{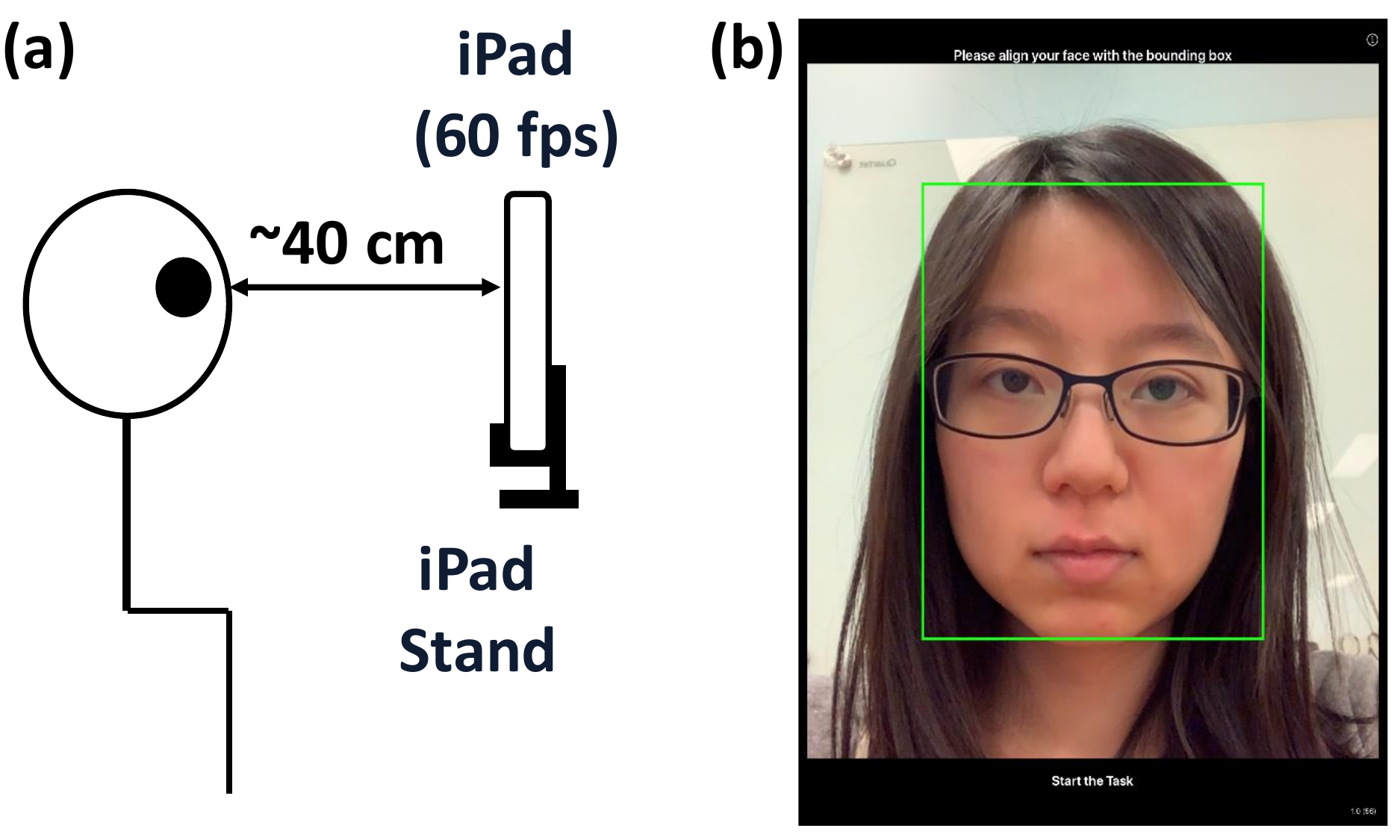}
     
      \caption{{(a) Recording setup; (b) before showing the task on the screen, the app displays the face of the subject with a bounding box. If the distance measurement from the camera to the subject's face is accessible (i.e. between 30 and 50 cm), the box will turn green. If the automatically detected ISO is greater than 1000, a warning will be shown to guide the subject to move to a better-illuminated place.} 
      }
      \vspace{-12pt}
       
      \label{fig:app-design}
  \end{figure}

To minimize the influence of environmental conditions on the quality of the recordings, we initially asked subjects to position the iPad at a distance of 30 to 50 cm. In subsequent releases for iPads with depth-sensing capability, the app senses the distance from the subject and provides visual feedback so the subject can position the iPad within the desired target distance (Fig.~\ref{fig:app-design}).
Besides distance, the app also guides the subject to position themselves in proper lighting conditions, and the subject will be asked to move to a brighter location if the automatically detected ISO is greater than 1000. 

When the subject is ready to perform the task, they start the recording, and a count-down will be displayed so the subject can begin to focus their attention on the pro-/anti-saccade task. The task will then be displayed while the frontal camera simultaneously records the subject's face. 
After the task is completed, a detailed set of data files is saved for each recording session, including the actual video recording, the timestamps of each recorded frame, the timestamps of each frame displayed on the screen, and a text file containing information about the recording system (iOS version, iPad generation), the distance of the iPad to the subject (when available), and the recorded ISO value at the beginning of the recording. 


To acquire accurate saccade latency measurements, it is crucial to synchronize the task display on the iPad screen and the recording from the iPad camera. We detailed and evaluated the synchronization in Appendix. By requiring the ISO to be less than 1000, we showed that we can bound the absolute synchronization error to be within 5 ms. 

\subsection{Task design} \label{sec:task-design}
\begin{figure}[b]
      \centering
      \includegraphics[width=\columnwidth]{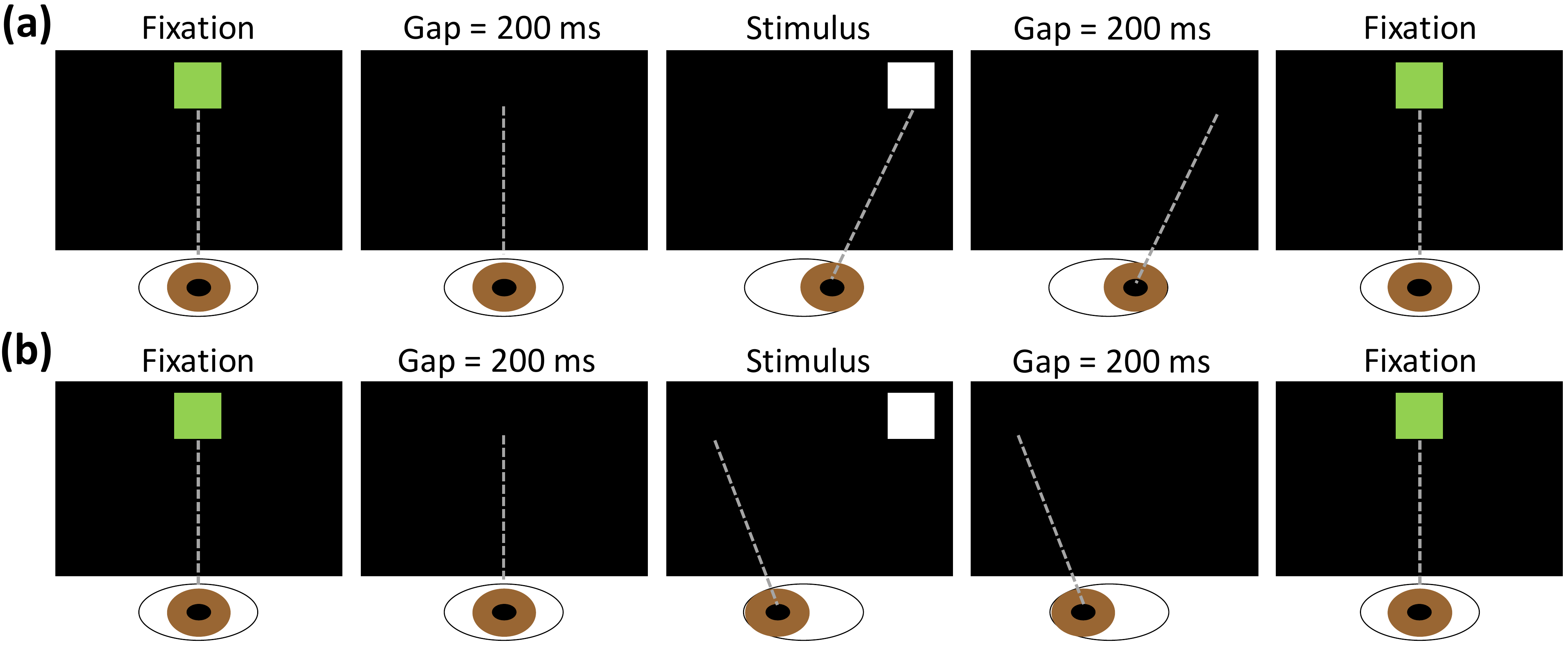}
     
      \caption{{(a) Pro-saccade task: Look toward the stimulus. (b) Anti-saccade task: Look away from the stimulus.} 
      }
       
      \label{fig:anti-task}
  \end{figure}

In this work, we implemented two commonly studied tasks in the literature, namely a gap-pro-saccade and a gap-anti-saccade task~\cite{Rivaud_2006,Garbutt_2008,Crawford_2005}. Both tasks start with a fixation period. During the fixation period (1 s), a fixation point (green square) is shown at the top center of the screen (as shown in Fig.~\ref{fig:anti-task}). We chose the fixation point to be presented at the top to prevent occlusion from the eyelids. Subjects were instructed to look at the fixation point during this period. The fixation is followed by a 200-ms gap period, where the fixation point disappears and the screen stays black. After the gap period, a stimulus (white square) is presented on either left or right side of the screen. If a subject is performing a pro-/anti-saccade task, the subject is instructed to move their eyes towards/away from the stimulus as quickly and accurately as possible. This stimulus period will last for 1.2 s and be followed by another 200-ms gap period. This sequence of ``fixation-gap-stimulus-gap'' will repeat for 20 or 40 times, with half of the stimuli presented to the right of the fixation point and half to the left in randomized order. 

%
%

\section{Methods} 
\label{sec:methods}

\subsection{Measurement Pipeline Requirements}
\begin{figure}
      \centering
      \includegraphics[width=\columnwidth]{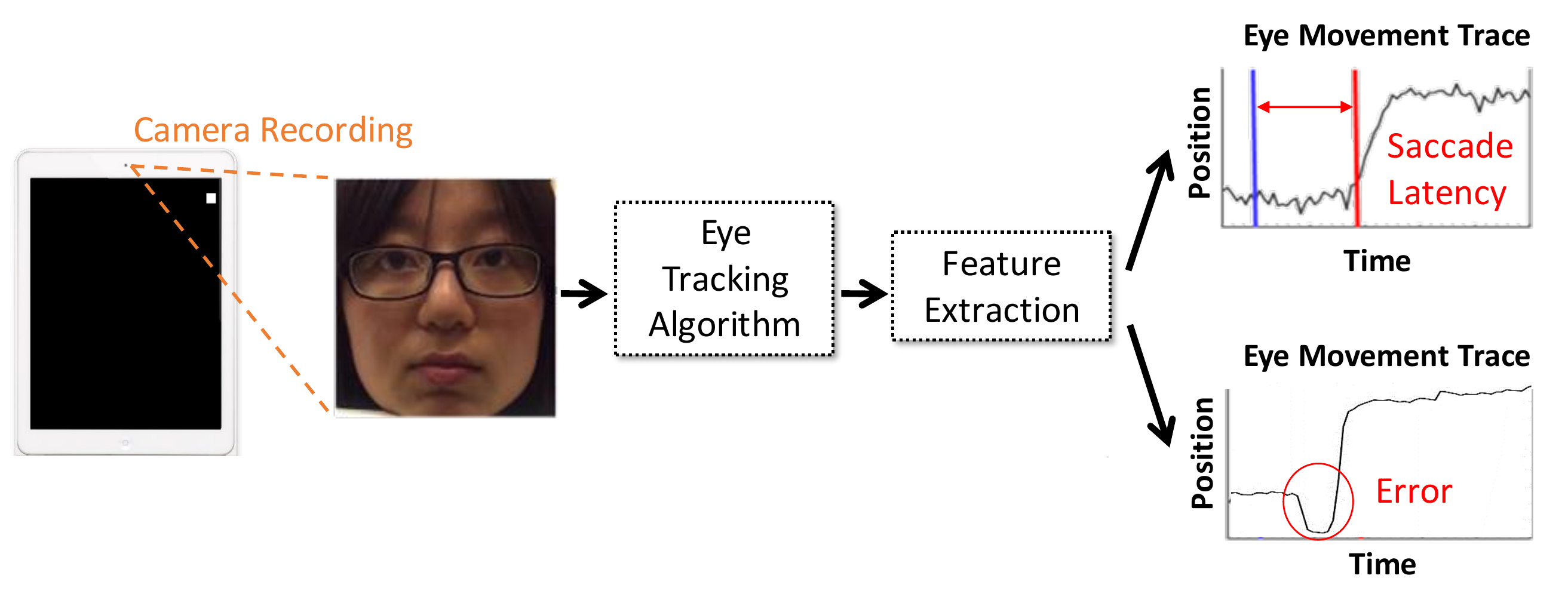}
     
      \caption{{The measurement pipeline includes the tablet-based video recording, an eye tracking algorithm, a saccade-latency measurement algorithm, and an error detection algorithm.} 
      }
       
      \label{fig:pipeline}
  \end{figure}

\begin{figure}
      \centering
      \includegraphics[width=\columnwidth]{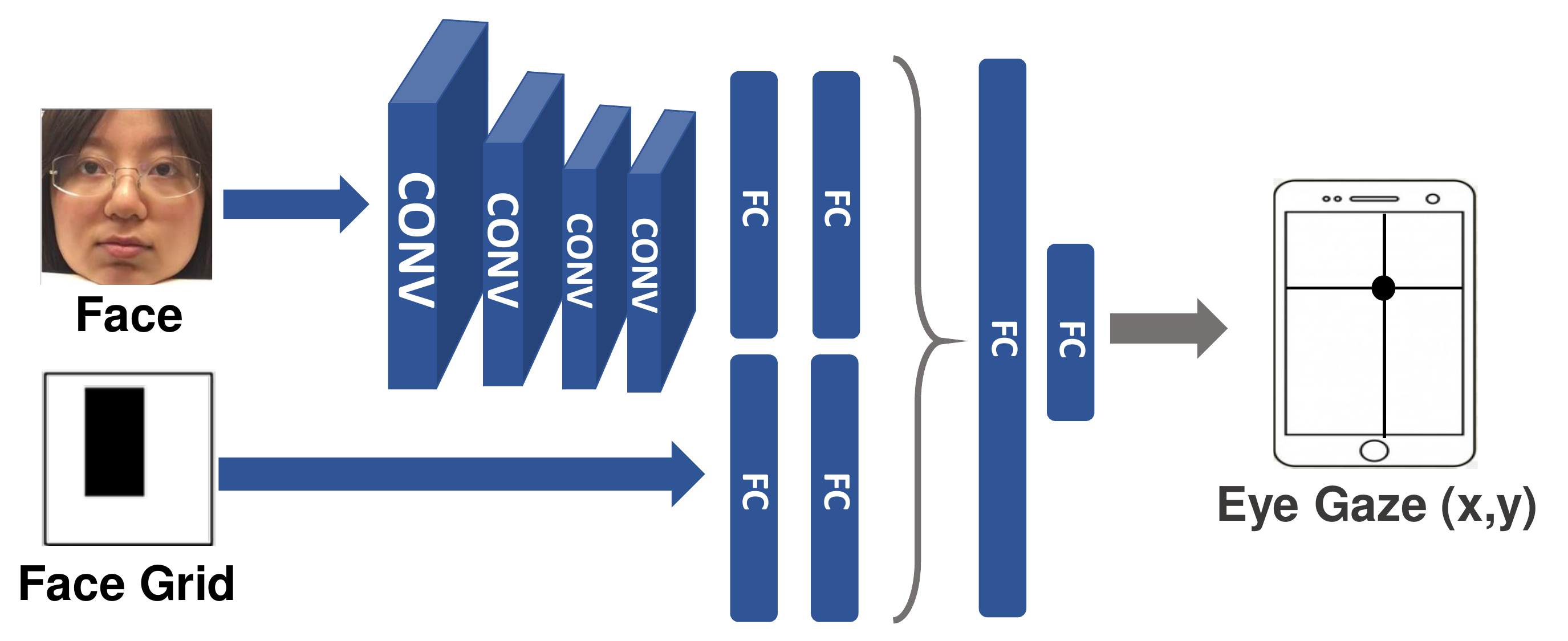}
     
      \caption{Convolutional neural network architecture used by iTracker-face. ``CONV'' stands for convolutional layers and ``FC'' stands for fully connected layers. The details of the architecture can be found in~\cite{Lai_2018}. 
      }
       
      \label{fig:iTracker}
  \end{figure}

In this section, we discuss our measurement pipeline as shown in Fig.~\ref{fig:pipeline}. Building on our prior work~\cite{Lai_2018, Saavedra_2018, lai_2020}, we used iTracker-face to estimate the gaze of the subject (Fig.~\ref{fig:iTracker}). The inputs to iTracker-face include the cropped face, as determined by the Viola-Jones algorithm~\cite{Viola_2001}, and a face grid indicating the face position. The outputs of iTracker-face are the (x, y)-coordinates of the estimated gaze position on the screen in the unit of centimeters. To attain our horizontal eye movement trace, we retain the x-coordinate of the gaze position across frames. 
As discussed in Section \ref{sec:materials}, we have synchronized the camera recording with the task display. We use the timestamps of the screen frames to acquire the time when each stimulus appears. With the stimulus presentation time and the eye movement trace, we can determine saccade latency and detect eye-movement errors.

In~\cite{lai_2020}, we measured saccade latency by fitting a hyperbolic tangent ($\tanh$) to a fixed window of the eye movement trace, from 100 ms before to 500 ms after the stimulus presentation, and determined the saccade onset as the time when the best model fit exceeded 3\% of the maximal saccade amplitude. Saccade latency was then computed as the time difference between the saccade onset and the time when the stimulus presented. A major benefit of using this model-based approach is that it provides an automated signal-quality quantification by means of the normalized root-mean-square error (NRMSE) between the model fit and the eye-position trace. We marked a trace as unusable if its NRMSE was greater than 0.1~\cite{lai_2020}. 
In addition, since the output of iTracker-face is in the unit of centimeters, we normalized the trace to the expected saccade amplitude using the best fit model. Since our saccade onset determination is scale invariant, the measured saccade latency is invariant of this normalization.

In the current study, we expanded upon our initial study cohort in~\cite{lai_2020} by specifically including self-reported healthy subjects across the adult age spectrum. Consequently, we observed a larger heterogeneity in saccadic eye-movement patterns that necessitated revisions to our previously established processing pipeline.
\begin{figure}[b]
      \centering
      \includegraphics[width=\columnwidth]{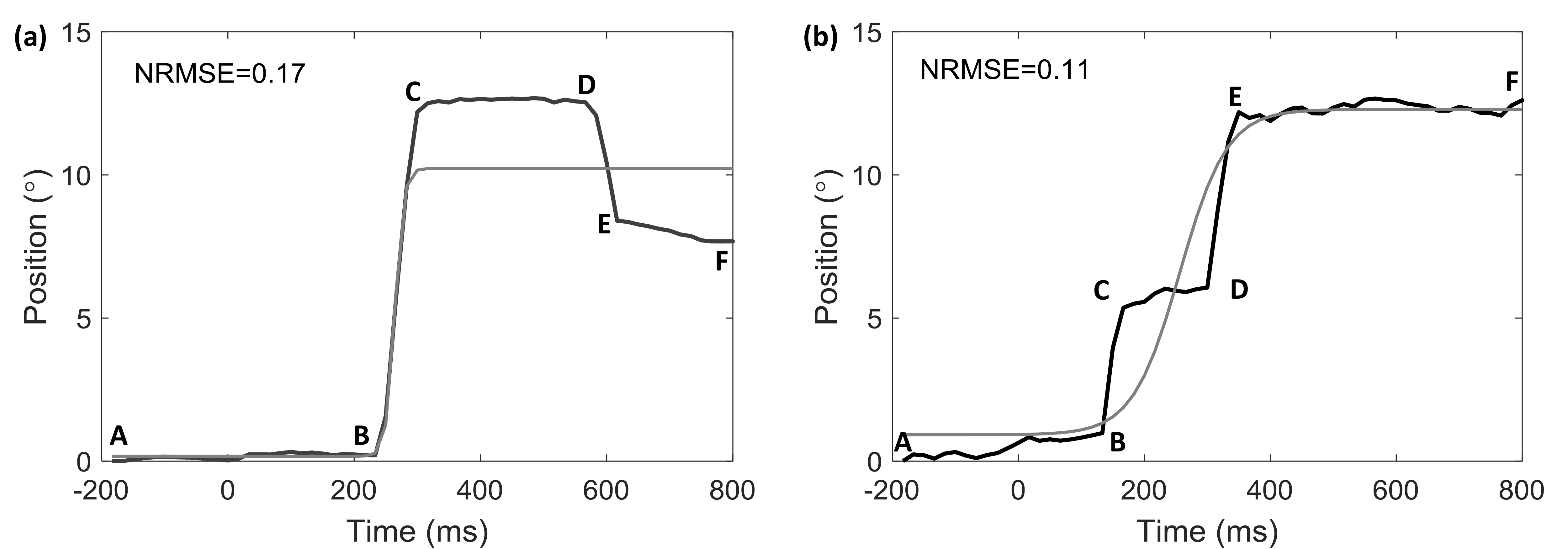}
     
      \caption{{Examples where $\tanh$ cannot be fitted to the entire trace: (a) gaze returning (b) hypometric saccade~\cite{Anderson_2013,Leigh_2015}. As one can see, to find the saccade latency, the window where we fit a $\tanh$ model should be from A to D.} 
      }
      \label{fig:hard-fit}
  \end{figure}
  
To allow for latency measurements from subjects with slower response times, we needed to increase the window of fit for the $\tanh$ model from 200 ms before to 800 ms after the stimulus presentation.
However, we noticed that by expanding the window, it is more likely to capture a subject's eye movements back toward the center position (Fig.~\ref{fig:hard-fit}a). Additionally, subjects may perform a series of hypometric saccades in which the initial saccadic movement does not reach the final position and a second saccade is made to correct for this undershoot (Fig.~\ref{fig:hard-fit}b). 
Correct identification of hypometric saccades is of relevance since an increased incidence of hypometric saccades is associated with certain neurodegenerative pathologies~\cite{Mosiman_2005,Anderson_2013}.
The single $\tanh$ model cannot fit well to these traces if we use a fixed window to determine latency values. To determine saccade latency, we need to allow for an adaptive window of fit for the $\tanh$ model to identify the initial saccadic movement to be fitted. 
We also note that we cannot convert the unit of the eye-movement trace from centimeter to degree using the best fit $\tanh$ model. As a result, we needed to revise our method to normalize the trace.

\subsection{Saccade Normalization} \label{sec:saccade-normalization}

We make three assumptions to convert the unit of the eye-movement traces from the iTracker-face generated centimeter to degree. First, we assume that subjects were looking at the fixation point during the fixation period. Second, we assume that subjects did not overshoot their gaze. Finally, we assume that during the stimulus period, subjects either (a) did not move their eyes at all, (b) gazed at the stimulus, or (c) gazed at the opposite position of the stimulus.

With these assumptions, we normalize the trace as follows. First, to simplify the algorithms, we flip the trace if needed so that positive excursions correspond to eye movements in the correct direction. We then smooth the eye-movement trace with a Savitzky-Golay filter~\cite{Nystrom_2010,Savitzky_1964} (of order 3 and frame length 5) to make the final normalization more robust to noise. Subsequently, we determine two reference points to scale and shift the eye-movement trace. 
Our first reference point is set as the starting gaze position of a trace, that is 200 ms before the stimulus presentation. With the second assumption, our second reference point is either the maximum or the minimum value of the smoothed trace, depending on whether the subject makes a correct saccade, a corrected error, or an uncorrected error. Scaling and shifting coefficients can be found by shifting the first reference point to zero degree and scaling the second to either the final expected amplitude (12.7 degrees) or the negative amplitude ($-$12.7 degrees). 

More precisely, we consider three scenarios. Operating on the output of iTracker-face, if the difference between the maximum value and the starting gaze position is greater than $0.2$ cm, we assume that the subjects have made a correct saccade or a corrected error, and we scale the second reference point to the positive expected amplitude value. If the difference between the maximum value and the starting gaze position is smaller than $0.2$ cm but the absolute difference between the minimum value and the starting gaze position is greater than $0.2$ cm, we assume that the subjects have made an uncorrected error and we scale the second reference point to the negative expected amplitude value. If neither of these criteria is met, we assume that the subjects have made only subtle eye movements or that the eyes were occluded. 

In the first two scenarios, we find the scaling and shifting coefficients from the smoothed trace and normalize the original trace using these coefficients. One key observation of this normalization is that after normalization, traces with the same shape will become identical. This characteristic ensures that if the saccade-latency measurement algorithm and the error-detection algorithm are designed using this normalized trace, the algorithms will be scale-and-shift-invariant. That is, eye movement features are measured only based on the shape of a trace.
In the third scenario, we noticed on visual inspection of the video recordings that the sizes of the eye movement were often comparable with noise and subtle head movement. To account for such observations, we label such traces as ``LS'' (Low Signal) to acknowledge the fact that we are uncertain whether there is an actual eye movement even by visualizing the original videos. Traces labeled LS will be excluded from the saccade-latency measurement and the error-detection algorithm.

\subsection{Adaptive Windowing and Saccade Latency Measurement} \label{sec:saccade-latency}

With the normalization, we next describe how we updated the window of fit for saccade latency measurement. Returning to the examples in Fig.~\ref{fig:hard-fit}, during the period from A to B and C to D, the subject's eyes are fixated. During the period from B to C, the subject performed a correct saccade, in the sense that the eyes moved in the correct direction. As a result, the proper window of fit is the first sequence of fixation, directionally correct eye movement, and fixation. This period can be identified using the velocity of the gaze. We estimate the velocity of the gaze by computing the first-order derivative of the Savitzky-Golay filtered trace to avoid amplifying high-frequency noise.

\begin{figure}
      \centering
      \includegraphics[width=\columnwidth]{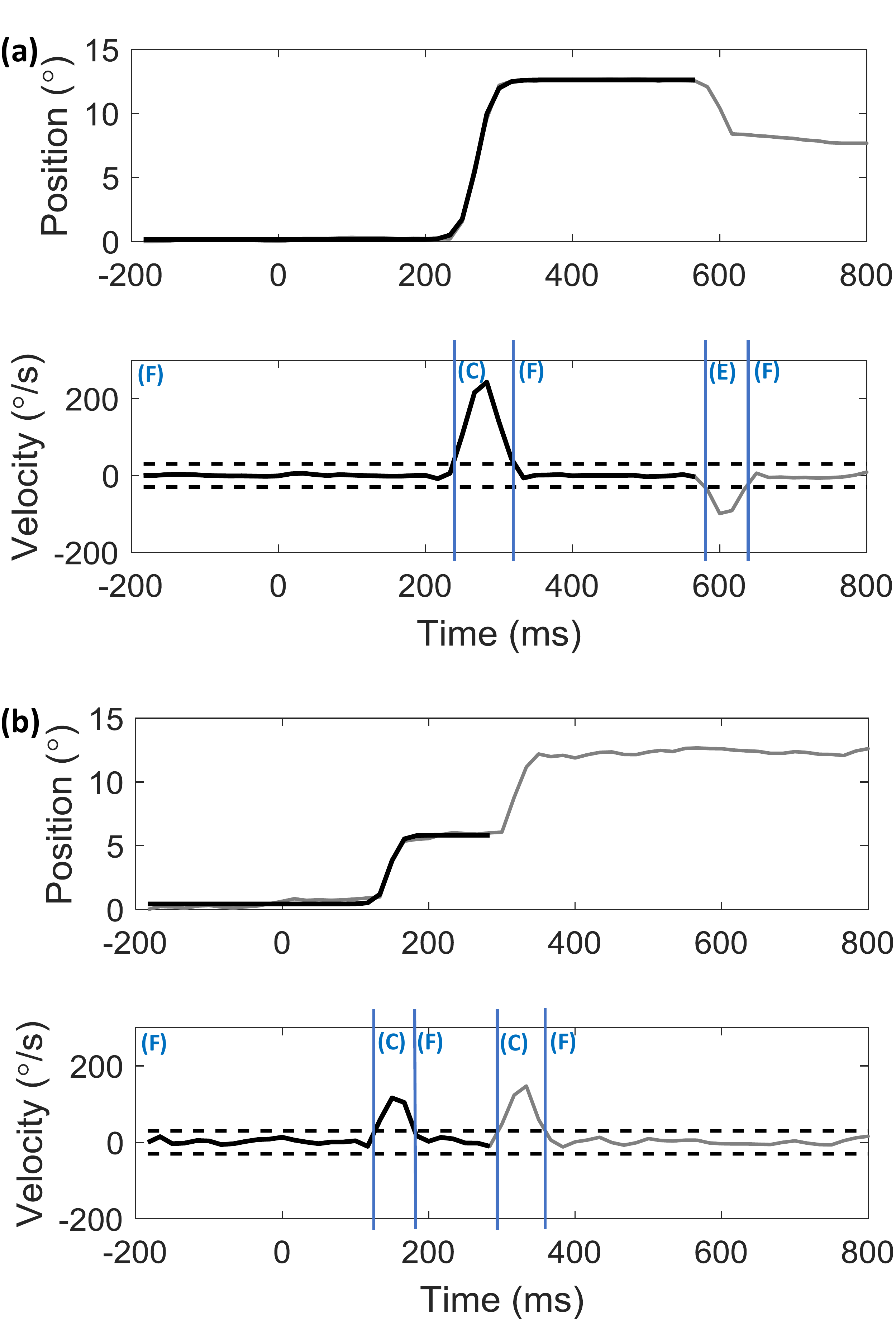}
     
      \caption{{Tanh fitting example: (a) gaze returning (b) hypometric saccade. The top panels show the eye movement traces obtained from iTracker-face after normalization. The dark lines show the fitted hyperbolic tangent models. The bottom panels show the velocity of the eye movements and the velocity threshold (the dash lines). With such a threshold, we label different parts of the trace as fixation (F), correct saccade (C), or error saccade (E). The window of fit is chosen as the first ``fixation(F)-correct saccade(C)-fixation(F)'' period that crosses a third of the amplitude.} 
      }
       
      \label{fig:new-tanh}
  \end{figure}

We then classify a sequence of time instances as a correct saccade period if the velocity values cross 30 degrees/s, as an incorrect saccade period if the velocity values cross $-$30 degrees/s, and as a fixation period otherwise. When there are more than one correct saccade periods, we will fit our model to the one that first crosses a third of the amplitude. Fig. \ref{fig:new-tanh} shows that by choosing the window of fit to be the period associated with the sequence of fixation, directionally correct eye movement, and fixation, we can fit the $\tanh$ model to traces with multiple transitions and measure their saccade latencies. We compared the previously described fixed-window approach with the adaptive-window approach and observed that the proportion of saccades with a NRMSE $>$ 0.1 dropped from 17\% to 3\% with the adaptive-window approach. Hence, by moving to the adaptive-window approach, we were able to compute significantly more latencies with this improved saccade-latency measurement algorithm.

\subsection{Error Detection} \label{sec:error}

In the clinical literature, a directional error is defined as an initial eye movement in the wrong direction~\cite{Rivaud_2006}. Manual annotation is often involved in the determination of these errors~\cite{Mosiman_2005,Crawford_2005}. Because such clinical studies have traditionally relied on specialized environments and eye-tracking equipment, including use of chinrests, infrared illumination, and research-grade cameras, there were usually comparatively few traces collected per subject and the traces tended to be clean. As a result, manual annotation of traces is possible in these cases. In contrast, to enable collection of large amounts of data, we use consumer-grade cameras and do not use a chinrest. As a result, we obtained significantly many more traces, though some were affected by glares or head movements. Our goal is thus to reject poor recordings and develop an accurate and robust error detection algorithm.

\begin{figure}
      \centering
      \includegraphics[width=\columnwidth]{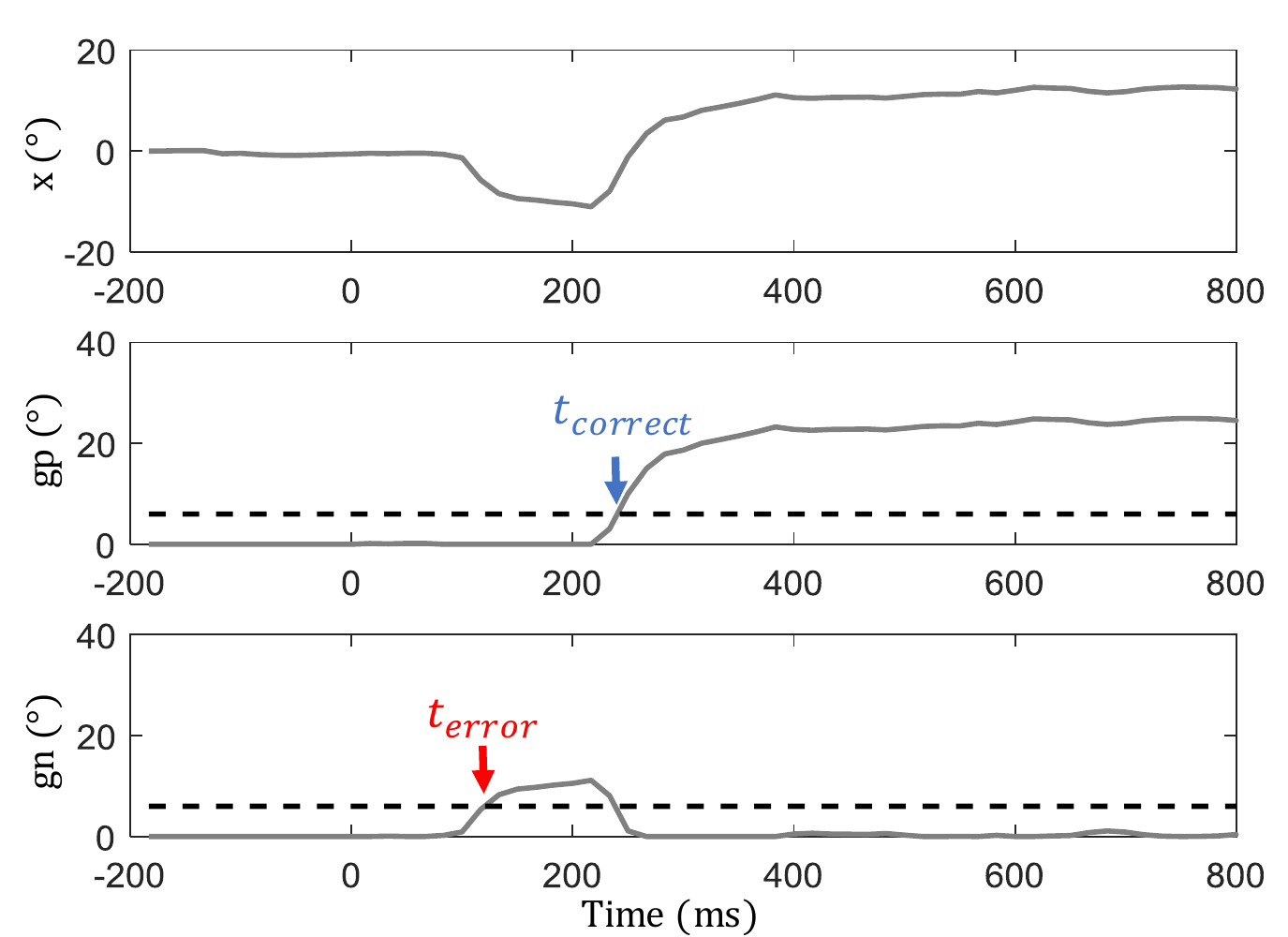}
     
      \caption{{Error detection example. The top panel shows the x coordinates of the iTracker-face output over time ($x_t$). The middle and the bottom panel show $gp_t$ and $gn_t$. The dashed line indicates the threshold $T$. When $gp_t$ and $gn_t$ cross the threshold $T$, $t_{correct}$ and $t_{error}$ are detected, respectively. In this case, since $0<t_{error}<t_{correct}$, an error is detected.} 
      }
       
      \label{fig:detect-error}
  \end{figure}

As mentioned in Section \ref{sec:saccade-latency}, we exclude the traces labeled LS, since we cannot distinguish between saccadic eye movements and noise/head movements. Out of the remaining traces, we noticed that a typical error trace shows a period of fixation followed by a directionally incorrect eye movement (as shown in the top panel of Fig. \ref{fig:detect-error}). Since our goal is to detect such a change, we developed our algorithm based on the change detection literature~\cite{Gustafsson_2001}. In particular, we extended the cumulative sum (CUSUM) algorithm~\cite{Page_1954} for our purposes.

%
%

%
%

We first assume that our measured eye movement trace $x_t$ at time $t$ is composed of an eye movement $\theta_t$ and an additive measurement noise $\epsilon_t$. We then use a recursive least square filter to estimate the eye movement $\hat{\theta}_t$ according to
\begin{equation}
\hat{\theta}_t=\lambda\hat{\theta}_{t-1}+(1-\lambda) x_t.
\end{equation}
The residual error then becomes $\hat{\epsilon}_t=x_t-\hat{\theta}_t$. 
If there is neither a positive trend nor a negative trend in $x_t$, $\hat{\epsilon}$ will be centered around zero. As a result, when we consider the cumulative sum of the residual error $s_t=s_{t-1}+\hat{\epsilon}_t$, $s_t$ will be centered around zero as well. However, if there is a negative trend in $x_t$ as shown in Fig. \ref{fig:detect-error}, $s_t$ will become progressively more negative. We can then use a threshold to determine whether $s_t$ is sufficiently negative such that $\epsilon_t$ is unlikely to just represent additive measurement noise. 

 To distinguish between correct saccades and incorrect saccades, we define two separate variables for $s_t$:  $gn_t=\max\{gn_{t-1}-\hat{\epsilon}_t,0\}$ and $gp_t=\max\{gp_{t-1}+\hat{\epsilon}_t,0\}$. That is, $gn_t$ accumulates negative trends and $gp_t$ accumulates positive trends. As a result, when $gn_t$ and $gp_t$ cross the pre-determined threshold, we detect an incorrect and a correct saccade, respectively. To apply the definition of a directional error as an initial eye movement towards the wrong direction, we detect an error if $gn_t$ crosses the pre-determined threshold after $0$ ms and before $gp_t$ crosses the pre-determined threshold.

Here, we chose to scale the threshold with respect to the estimated (corrected) saccade amplitude. We notice that if there is no error in a trace, $gn_t$ will be around zero while $gp_t$ will approximate the amplitude of the saccade (Fig. \ref{fig:detect-error}). When there is an error, $gp_t$ will approximate the amplitude of the corrected saccade. On the other hand, when there is an uncorrected directional error, $gn_t$ will approximate the amplitude of the saccade. As a result, we approximate the (corrected) saccade amplitude by $max_t \{gp_t,gn_t\}$. We further observe that if the saccade amplitude before the normalization is sufficiently large, the saccade will be less affected by head movement and noise. Thus, we can consider lowering the threshold to detect smaller errors. On the other hand, if the original saccade amplitude is closer to the size of the head movement and noise, the threshold needs to be sufficiently large to avoid artifacts from being detected. Recall that in Section \ref{sec:saccade-latency}, we scale the trace and shift it to normalize it from centimeters to degrees. We can use the scaling coefficient (denoted as $B$ in \textbf{Algorithm}) as a metric to evaluate the size of the original saccade amplitude. If $B$ is small ($<8$), it means that the original amplitude is large and the threshold could be smaller. If $B$ is large ($\ge 8$), we will use a fixed threshold. Here the value $8$ can be considered as a hyperparameter that we can tune.
The final threshold is $\max_t \{gp_t,gn_t\}\cdot \min \{B,8\} \cdot T$.  The complete algorithm is shown in \textbf{Algorithm}. 

\begin{algorithm}
\SetAlgoLined
\SetKwInOut{Input}{input}
\SetKwInOut{Output}{output}
\SetKwInOut{Parameter}{parameter}
\Input{$x=[x_1,\ldots,x_N], B$, $x_1$ is chosen to be the first instance after the stimulus presentation, $B$ is the scaling coefficient in the saccade normalization}
\Output{$t_{error},t_{correct}$ (An error is only detected if the first element in $t_{error}$ is smaller than the first element in $t_{correct}$.)}
\Parameter{$\lambda,T$}
 \For{round=0:1}{
 $\hat{\theta}=x_1,t_{error}=[], gn=[0], gp=[0]$\;
 \For{t=2:N}{
  $\hat{\theta}=\lambda \hat{\theta}+(1-\lambda)x[t]$\;
  $\hat{\epsilon}=x[t]-\hat{\theta}$\;
  $gn.append(\max\{gn[t-1]-\hat{\epsilon},0\})$\;
  $gp.append(\max\{gn[t-1]+\hat{\epsilon},0\})$\;
  \If{round==1}{
    \If{$gn[t]>A\cdot T$}{
       $t_{error}.append(t)$\;
       $gn[t]=0$\;
       $\hat{\theta}=x[t]$\;
    }
    \If{$gp[t]>A\cdot T$}{
       $t_{correct}.append(t)$\;
       $gp[t]=0$\;
       $\hat{\theta}=x[t]$\;
    }
   }
 }
 $A=\min\{8,B\}\cdot \max\{gp,gn\}$\;
 }
 \caption{Error Detection}\label{algo:error}
\end{algorithm}

\begin{figure}
      \centering
      \includegraphics[width=0.8\columnwidth]{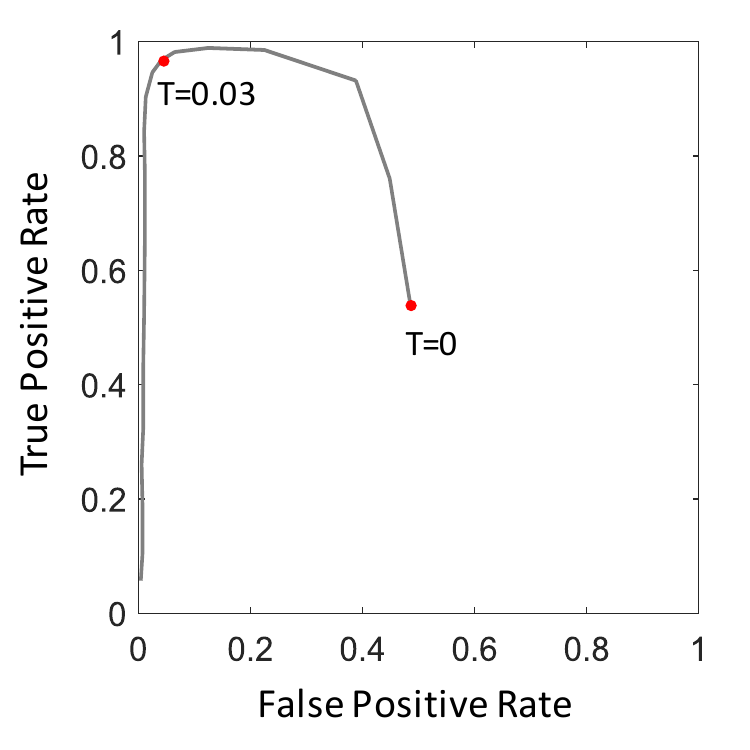}
     
      \caption{{The true positive rate and the false positive rate as we increased the error detection threshold $T$ from 0 to 0.1. We chose $T=0.03$ as our final threshold to achieve a sensitivity of 0.97 and a specificity of 0.97.} 
      }
       
      \label{fig:ROC}
  \end{figure}

To determine the threshold $T$, we asked four subjects to perform six anti-saccade tasks of 40 stimuli each. Two expert annotators reviewed the videos and annotated the directional errors. Out of the $4\cdot 6\cdot 40=960$ saccadic eye movements, there were only two disagreements between the annotators which were resolved after these two disagreements were reviewed together. 
With the annotated data set at hand, we swept the threshold $T$ and determined the true positive and false positive rates for detecting a directional error (Fig. \ref{fig:ROC}). When the threshold is lower than the noise level, $gp_t$ and $gn_t$ may cross the threshold due to noise rather than a saccadic eye movement. That is, $gp_t$ may be equally likely to cross the threshold as $gn_t$. Recall that we only detect a trace as an error if $gn_t$ crosses the threshold before $gp_t$. As $T$ goes to zero, the true positive rate and the false positive rate go to 0.5. On the other hand, if the threshold is too large, the amplitude of an incorrect saccade may be smaller than the threshold and the error may not be detected.  When $T$ is larger than the noise level but smaller than the amplitude of an error, we can get high sensitivity and specificity. By choosing $T=0.03$, we can achieve a sensitivity of 0.97 and a specificity of 0.97 for detecting a directional error.

\subsection{Error Rate Definition}
In the clinical literature, error rate is often defined as the proportion of errors, though it is not usually discussed whether noisy traces are excluded from such calculation. Given the use of  special-purpose equipment and optimized environmental conditions in clinical research studies, such recordings may have very few noisy traces. Without a chinrest and a controlled laboratory setup, we obtained more noisy traces. 
We carefully identified the causes of these noisy traces: glares, head movements, eyelids drooping. Many of these causes could be reduced with more careful instruction. However, even with careful instruction, it is hard to eliminate all these causes, due to the nature of the much more relaxed and varying recording environment and the large number of recordings. As a result, it is important to define an error rate that takes these noisy traces into consideration.

An eye movement was either declared a correct saccade (dC), declared an error (dE), or labeled low signal (LS). If we define the error rate as the proportion of errors out of all the traces, we might significantly underestimate the error rate in records with a lot of eye movements in the LS category. A better approach might be to define the error rate in a recording as \#dE/(\#traces-\#LS), as explained in Eq. (\ref{eq:error-rate}). The question then arises under which conditions the error rate so defined approximates the (empirical) probability of an error. 

Under the assumption that 
\begin{itemize}
    \item $P(dE|C)\approx 0, P(dC|E)\approx 0$,
    \item $P(LS|E)\approx P(LS|C)$,
\end{itemize}
where E denotes errors and C denotes correct saccades, we can express the error rate as
\begin{equation}
    \begin{aligned}
&\frac{P(dE)}{1-P(LS)}\\
=&\frac{P(dE|E)P(E)+P(dE|C)P(C)}{1-P(LS|E)P(E)-P(LS|C)P(C)}\\
\approx & \frac{P(dE|E)P(E)}{P(E)[1-P(LS|E)]+P(C)[1-P(LS|C)]}\\
\approx & \frac{[1-P(LS|E)]P(E)}{[1-P(LS|E)]P(E)+[1-P(LS|C)]P(C)}\\
\approx & \frac{P(E)}{P(E)+P(C)}\\
= & P(E)
\end{aligned} \label{eq:error-rate}
\end{equation}
where we made use of the fact that a trace is either an error or a correct saccade, i.e. $P(E)+P(C)=1$. 
The first assumption states that the false positive and the false negative are essentially zero. As discussed in Section~\ref{sec:error}, our error detection algorithm achieved a sensitivity of 0.97 and a specificity of 0.97. Therefore, the first two assumptions are indeed met. The second assumption states that a correct saccade is equally likely to be declared LS as an error saccade. Since our determination of LS is simply based on the size of the trace, this condition is met as well. Therefore, it is reasonable to define the error rate as \#dE/(\#traces-\#LS) as an estimate of the (empirical) probability of an error. 
\section{Data Analysis} \label{sec:dataanalysis}

\begin{figure}
      \centering
      \includegraphics[width=\columnwidth]{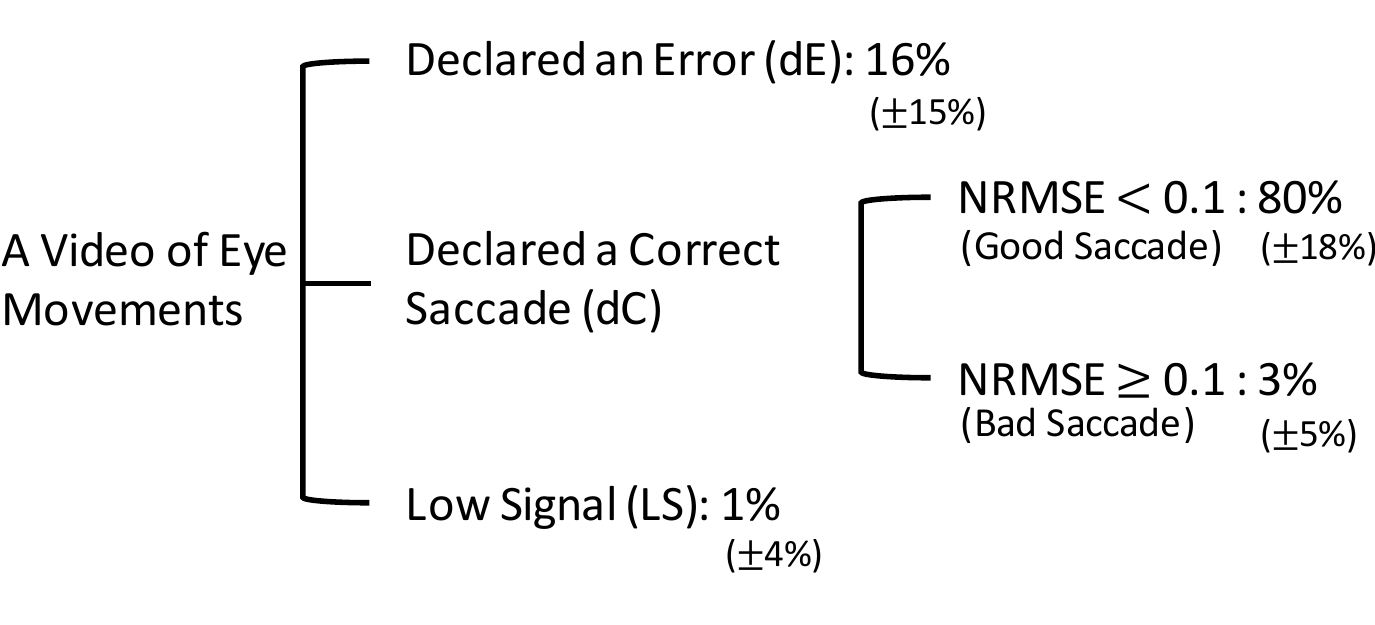}
     
      \caption{{Breakdown of saccades into error saccades, good saccades, bad saccades, and ``LS'' (low signal).} 
      }

      \label{fig:Bifurcation}
  \end{figure}

With our system, we have collected 6823 videos and 236900 eye movements from 80 subjects across the adult age spectrum.
With the saccade latency and error determinations, we labeled the traces as in Fig.~\ref{fig:Bifurcation}. We observe that in videos with a substantial number of LSs, subjects' eyes were often partially occluded due to eyelid droop. Videos with a large number of bad saccades tend to contain more head movements. As a result, the number of LSs and bad saccades indicates whether a subject recorded themselves properly. We therefore discard a video if more than half of the saccades are LSs or bad saccades. After discarding the videos with too many LSs and bad saccades, we retained 6787 videos and 235520 eye movements from 80 subjects. Out of the remaining videos, we calculated the mean (standard deviation) of the proportions of each label in a video. There are 1\% (4\%) of LSs and 3\% (5\%) of bad saccades. That is, on average, 96\% of the saccades are good saccades or declared errors. 

\begin{figure}
      \centering
      \includegraphics[width=\columnwidth]{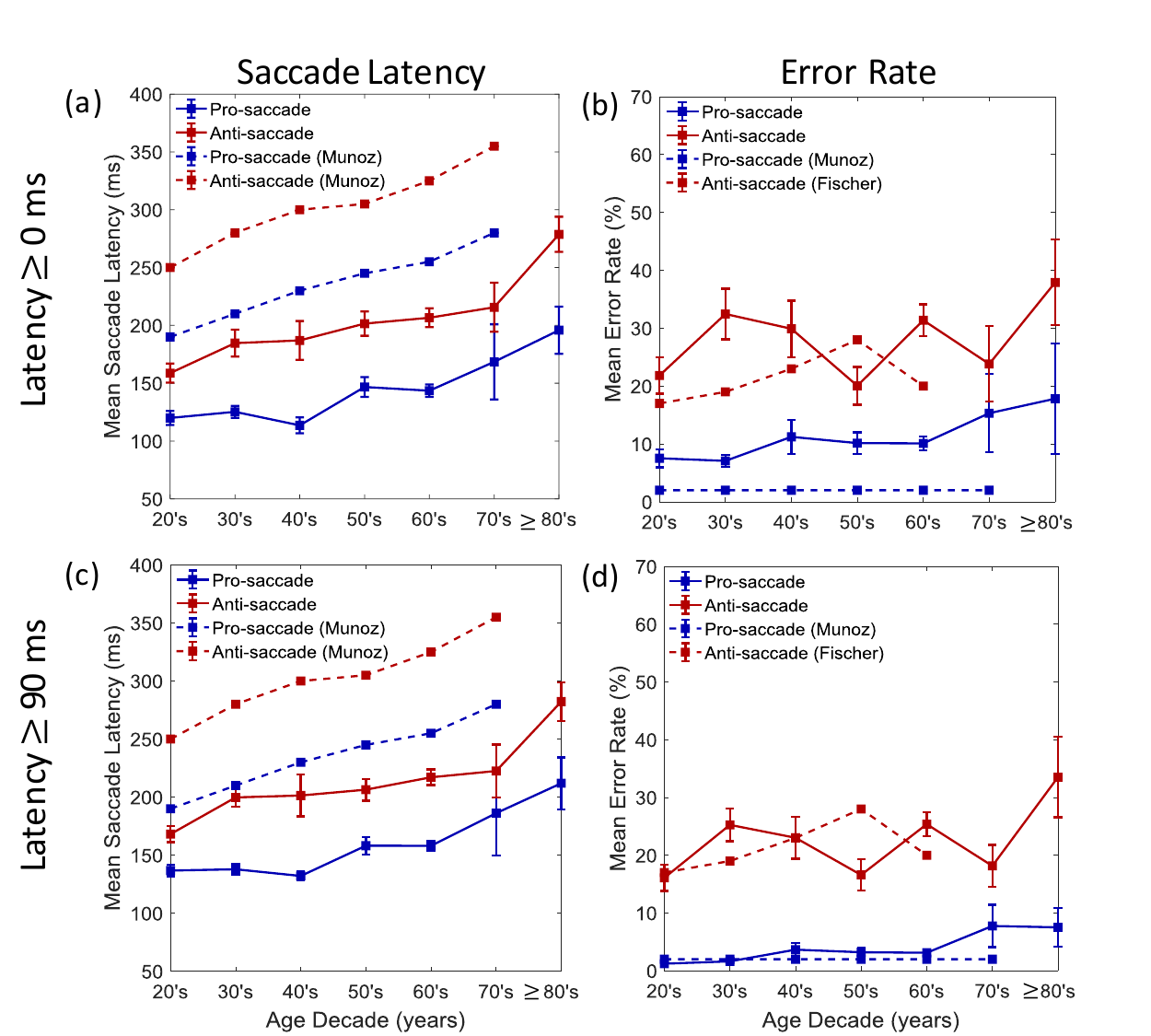}
     
      \caption{{Eye movement features as a function of age with saccades $>0$ ms: (a) mean saccade latency (b) mean error rate, and with saccades $>90$ ms: (c) mean saccade latency (d) mean error rate. The bars showed one standard error.} 
      }
      \vspace{-12pt}
      \label{fig:Analysis_0_90}
  \end{figure}

With these data, we can analyze the responses of eye movement features in different age groups (Fig.~\ref{fig:subject}). This is important because it gives us a baseline when we compare the results with data from patients. We calculate the mean saccade latency and error rate for each individual and then compute the mean and standard error of the individual mean saccade latencies and error rates per age group. As a result, the mean of an age group is not biased towards those subjects who provided more recordings. To evaluate the correlation between age and eye movement features, we compared our result with~\cite{Munoz_1998, Fischer_1997}, where data were collected from specialized equipment (DC electrooculography with a head rest) in a controlled environment. We notice that~\cite{Munoz_1998} defined an anticipatory saccade as any saccade (including errors) with latency $<$ 90 ms. To evaluate how changing this threshold may affect the result, we show the data with and without this anticipation threshold (Fig.~\ref{fig:Analysis_0_90}).

Several observations are worth noting. First, since anti-saccade tasks are more complex and require more cognitive processing~\cite{Crawford_2005,Mosiman_2005,Shafiq_2003}, the mean anti-saccade latency and anti-saccade error rate in every age group is larger than the corresponding mean pro-saccade latency and pro-saccade error rate. Moreover, we see that the saccade latency is positively correlated with age, whereas the correlation between error rate and age is not significant. These observations are in agreement with the data by Mu\~noz et al.~\cite{Munoz_1998}, though the actual saccade latency values in our study tend to be lower than those reported by Mu\~noz et al. Our results suggest that our measurement system and processing pipeline can identify similar trends as shown in the clinical literature. 
Another observation is that the definition of anticipatory saccades affects the measured pro-saccade latency and error rate. On one hand, this observation is reasonable, since pro-saccade tasks are much easier to perform and errors tend to be caused by anticipation. On the other hand, while there is no consistent definition of anticipatory saccades in the literature, our observation highlights that they should be carefully defined.

\begin{figure*}
      \centering
      \includegraphics[width=\textwidth]{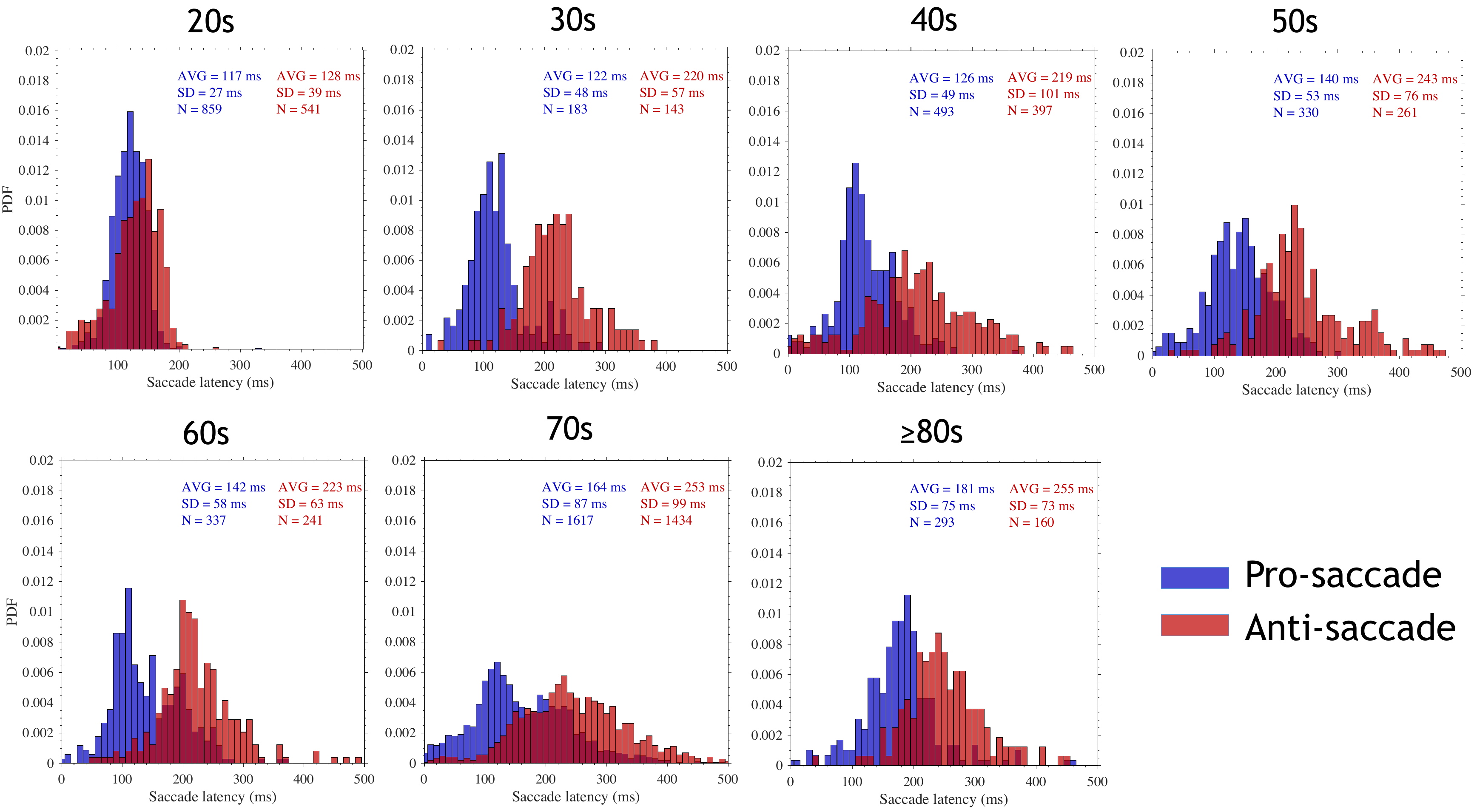}
     
      \caption{{Representative normalized distributions, shown as probability density functions (PDFs), of pro-saccade (blue) and anti-saccade (red) latencies for each decade in age of the study population. Subjects whose mean pro-saccade latency is the median of the corresponding age group were chosen to represent each group. No censoring was applied to eliminate anticipatory saccades. AVG: average latency; SD: standard deviation; N: number of eye movements.} 
      }
      \vspace{-12pt}
       
      \label{fig:distributions}
  \end{figure*}
  
In addition, with the accessibility to sizable data, we can study individual distributions, instead of only reporting the population mean as in most clinical literature. We analyzed the mean pro-saccade latency of each subject in seven age groups and chose from each age group the subject with the median mean pro-saccade latency as the representative subject. In Fig.~\ref{fig:distributions}, we showed example saccade latency distributions of these representative subjects. We observe that there are significant intra- and inter-subject variations in saccade latency across our study cohort, which suggests that aggregated results may lose the information encoded in individual distributions.
\section{Discussion} \label{sec:discussion}
The neural circuits involved in generating eye movements can be affected by neurodegenerative diseases. In particular, pro-/anti-saccade latency and error rates have been shown in the clinical literature to be significantly different between healthy subjects and patients with certain neurodegenerative conditions such as Alzheimer's disease and Parkinson's disease~\cite{Crawford_2005,Mosiman_2005,Shafiq_2003}. Thus, such eye movement features may be promising candidates for tracking disease progression.
However, these features are commonly measured in special, somewhat artificial environments and with special-purpose infrared-illuminated cameras, which limits broad accessibility and repeat measurements to track neurodegenerative disease progression longitudinally. In this work, we present, validate, and use an iOS application to enable such data collection. Additionally, we present algorithms for measuring saccade latency, determining directional errors, and calculating error rate that takes into account the possibility that it might not always be possible to determine the eye movement from home-based recordings.

\noindent\textit{Recording setup} \\
In our previous work, we showed that instead of a special-purpose camera, we can measure saccade latency using a smartphone camera. The recording setup, nevertheless, required a laptop to display the task, a screen synchronized with the laptop to be placed behind the subject, and a researcher to record both the subject's eye movement and the synchronized screen using the back camera of an iPhone. Due to these requirements, the recording setup was not sufficiently flexible for a subject to take recordings on their own in their homes or offices, which limits the possibility of using such a system to flexibly and ubiquitously monitor neurocognitive decline or disease progression. In this work, we designed an iOS app to record a subject with the frontal camera of an iPad while the subject is following a task shown on the screen. There are two challenges to achieve this goal.

First, unlike in the clinical setup and in our previous work where an expert researcher takes recordings of a subject, our app needs to guide the subject to record themselves at a proper distance to the camera and in a well-lit environment. To resolve this first challenge, before recording a subject, the app displays the subject on the screen and guides the subject to align their face with a bounding box shown on the screen. With such guidance, most subjects were recorded at an appropriate distance. To ensure the environment is well-lit, the app also asks the subject to move to a better-illuminated environment if the measured ISO is greater than 1000. 

Second, the camera recording and the task displayed on the screen need to be well-synchronized to obtain accurate saccade latency. This can be challenging as most applications (e.g., video chatting) only require the synchronization error to be unnoticeable by a human (i.e., $<$ 80 ms). With careful app design and evaluation of the synchronization error, we show that we can restrict the absolute timing error to be within 5 ms, which is well within the standard deviation of a subject's saccade latency distribution.

\noindent\textit{Algorithm design} \\
In our previous work, we measured pro-saccade latency by combining a deep convolutional neural network for gaze estimation with a model-based approach for saccade onset determination that also provides automated signal-quality quantification and artifact rejection. Here, we also include an anti-saccade task and extend our measurement pipeline to measure the associated saccade latencies and error rates. 

Since eye movements are now recorded outside of a clinical environment, our first observation is that in cases where eye movements are too small in amplitude or when the eyes are occluded, the eye movement signals can be smaller than noise. In these cases, we cannot tell the direction of the eye movement either from the trace or from the original video. As a result, we cannot classify these traces into a correct or an erroneous eye movement and cannot determine the saccade onset. We show that we can identify these traces using the raw output of iTracker-face, label these traces as the "LS"s (low signal), and exclude them from the saccade latency measurement and error detection.

Our second observation is that, since we now implement both pro- and anti-saccade tasks and that anti-saccade latencies are usually larger than pro-saccade latencies, we need to increase the size of the window where we fit our $\tanh$ model. However, by doing so, we also increase the potential of including more than one saccade movement in the window. For example, subjects may make a hypometric saccade or return their gaze towards the center of the screen. Being able to measure saccade latency from these traces is crucial, especially when these eye movements indicate a certain phenotype. For instance, patients with Parkinson's disease may make more hypometric saccades~\cite{Anderson_2013,Mosiman_2005} than patients age-matched controls. Our previous saccade latency measurement algorithm cannot measure latencies from these traces since a $\tanh$ model with a fixed window cannot fit well on these traces. 
In this work, we show how we can find the appropriate windows of fit for these traces and thus enable saccade latency measurement. By doing so, we keep 96\% of the traces to be either a good saccade (the saccade with NRMSE $\le$ 0.1) or an error saccade, which is much more than 82\% of the traces to be either a good saccade or an error if we use a fixed window.

Our third observation is that, to detect directional error is the same as to detect a change in the negative direction in an eye-movement trace. We extend the CUSUM algorithm for this purpose and show that our error detection algorithm can achieve a sensitivity of 97\% and a specificity of 97\%. 
Our final observation is that, given the absence of infrared illumination, high-speed cameras, and chinrests, there may be more LSs and bad saccades (saccades with NRMSE $>$ 0.1) in recordings where the subject did not record themselves properly or had several head movements. If we still define the error rate as the proportion of errors out of all the saccades as in the clinical literature, we may underestimate the error rate. As a result, after discarding undesirable recordings (recordings with more than half of the saccades being LSs or bad saccades), we define the error rate as the proportion of errors excluding LSs and show that this definition is a reasonable approximation for the error rate used in the clinical literature. 

\noindent\textit{Age and eye movement features} \\
With the improvement in our measurement pipeline, we took 6823 recordings from 80 subjects ranging in age from 20 years to 92 years, a significantly larger number compared to our previous work -- around 500 recordings from 29 subjects mostly in their 20's and 30's, and most other work collected just a few recordings from each subject~\cite{Mosiman_2005,Yang_2013,Rivaud_2006}. Moreover, we have 43 subjects with multiple recording sessions compared to 11 subjects in our previous work. Even after discarding undesirable recordings, we retained 6787 recordings and 235520 eye movements from 80 subjects.

As in the literature, we observe that anti-saccade latency and error rate tend to be larger than pro-saccade latency and error rate, respectively. Across the age range, we also observe that saccade latency is positively correlated with age while a strong relationship between error rate and age is not apparent. This observation also matches the observation in prior work~\cite{Munoz_1998,Fischer_1997}. Although our saccade latency values are smaller than values reported in~\cite{Munoz_1998,Fischer_1997}, our values are within the range of latency values reported in the clinical literature~\cite{Rivaud_2006,Crawford_2005,Holden_2018,Bonnet_2014}. Several hypotheses can be made to explain why our values may be smaller. First, our recording setup is less constrained. As mentioned in~\cite{NAP_2018}, recording subjects in dedicated environments may affect a subject's cognitive awareness. Second, our subjects are mostly graduate students or professors. It is likely that education level may affect reaction time. We also have fewer subjects in the 70's and 80's than in other age brackets. While one of the three subjects in the 70's has latency values much closer to the values reported in the literature, two other subjects have smaller latency values.

We also observe that the definition of an anticipatory saccade may significantly affect the measured pro-saccade latency and error rate. While the definition is not consistent across the clinical literature, our observation suggests that a more careful investigation into the effect of picking a latency threshold for anticipatory saccages on mean saccade latency is warranted. Some investigations designed tasks to avoid anticipatory saccades~\cite{Boxer_2012,Hopf_2018}, for example, by randomizing the length of the fixation period or by including more positions where a stimulus can be presented. However, we suspect that these modifications may result in an increased error rate. Since we aim to design and validate our error detection algorithm in this work, we did not implement either of these modifications. Nevertheless, it is worth analyzing how these modifications may affect saccade latency and error rate.

Last but not least, we show that with multiple recordings from each subject, we can study individual saccade latency distributions, while most literature only reported population means. We observe that there is significant intra- and inter-subject variability in these distributions. This observation suggests that such distinctive differences within and across subjects is lost if we were only to report a single summary statistic (mean or median) for each subject or across each age group. Our pipeline fundamentally enables the collection and the analysis of a large number of measurements to characterize the distributional characteristics for each subject. 


\section{Conclusion}

In this work, we developed, validated, and deployed an app to allow for robust determination of pro- and anti-saccade latencies in a visual reaction task. Additionally, we extended our previously reported signal processing pipeline to automatically detect low-signal recordings that should not be further analyzed and also identified directionally erroneous eye movements. With this platform in place, we collected over 235,000 eye movements from 80 self-reported healthy volunteers ranging in age from 20 to 92 years, an order of magnitude more measurements than presented in our previous work. We observed that pro- and anti-saccade latency is positively correlated with age whereas the relationship between error rate and age is not significant. Moreover, we observed notable intra- and inter-subject variability across participants, which highlights the need to track eye-movement features in a personalized manner. By enabling app-based saccade latency measurements and error rate determination, our work paves the way to use these digital biomarkers to aid in the quantification of neurocognitive decline and possibly from the comfort of the patient’s home.  
\section*{Acknowledgments}

The authors would like to thank Mr. Peter Kamm for help with the development of the app.
\bibliographystyle{IEEEtran}
\bibliography{IEEEabrv,refs}
\appendix{}
\label{sec:appendix}

In this appendix, we detail how we bound the error associated with saccade latency determination using the app. 
The accuracy of the saccade latency determination depends on the accuracy with which the timing of two events can be determined with the app, namely 1) the times of first presentation of each stimulus (the ``stimulus timestamps'' $s_i$), and 2) the times associated with the frame-by-frame recording from the camera (the ``recording timestamps'' $t_j$). Our typical saccade task consisting of 40 individual pro-/anti-saccade stimuli. Hence, we obtain 40 stimulus timestamps (i.e. $i \in [1, ..., 40]$), whereas we record around $Z \approx 6575$ frames from the camera for each recording (i.e. $j \in [1, ..., Z]$). Both series of timestamps are obtained through function calls to the operating system.

If the timestamps $s_i$ and $t_j$ could be obtained to very high accuracy, the resulting error in the saccade onset determination would solely be due to the saccade onset determination algorithm. However, given that operating systems generally prioritize a host of housekeeping tasks, timing information obtained from the operating system tend to be affected by queued access to the processor clock.

\begin{figure}[b]
      \centering
      \includegraphics[width=\columnwidth]{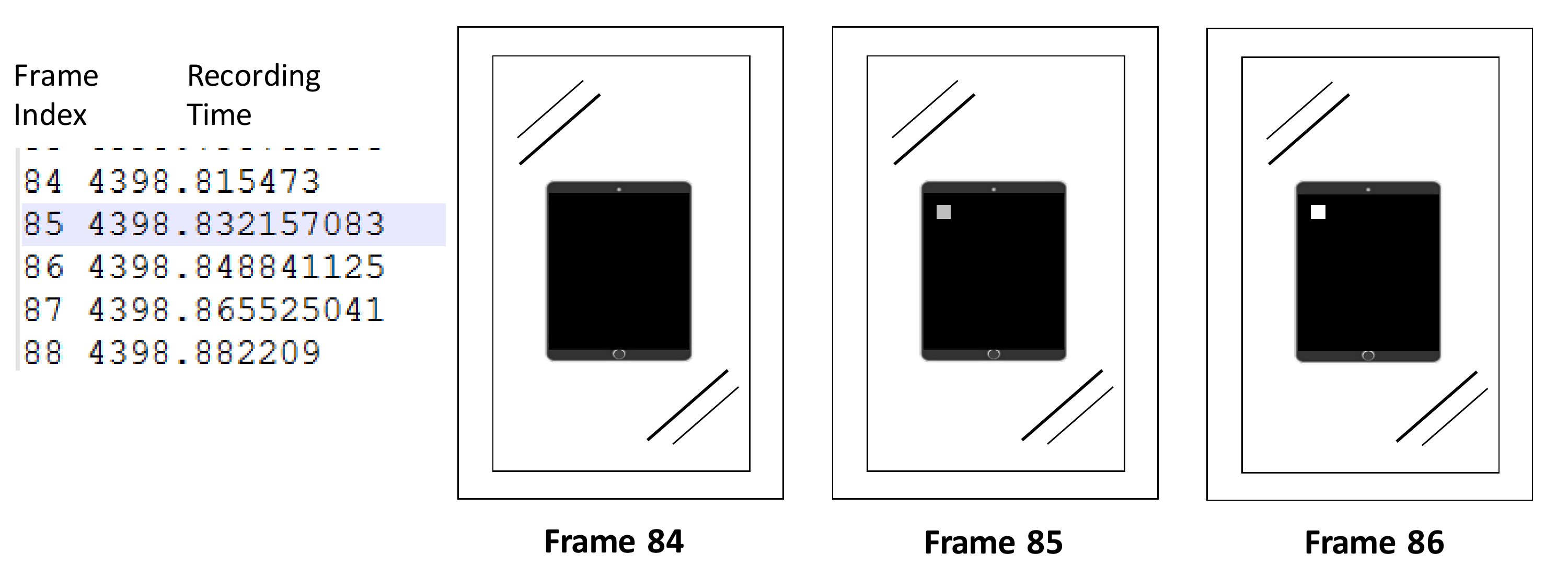}
     
      \caption{{Example for determining $r_i$, the time when the $i$-th stimulus appears. In this example, the first stimulus appears in recording frame 85 at $r_1=4398.8322$ s. } 
      }
       
      \label{fig:recording}
  \end{figure}
  
  \begin{figure}[b]
      \centering
      \includegraphics[width=\columnwidth]{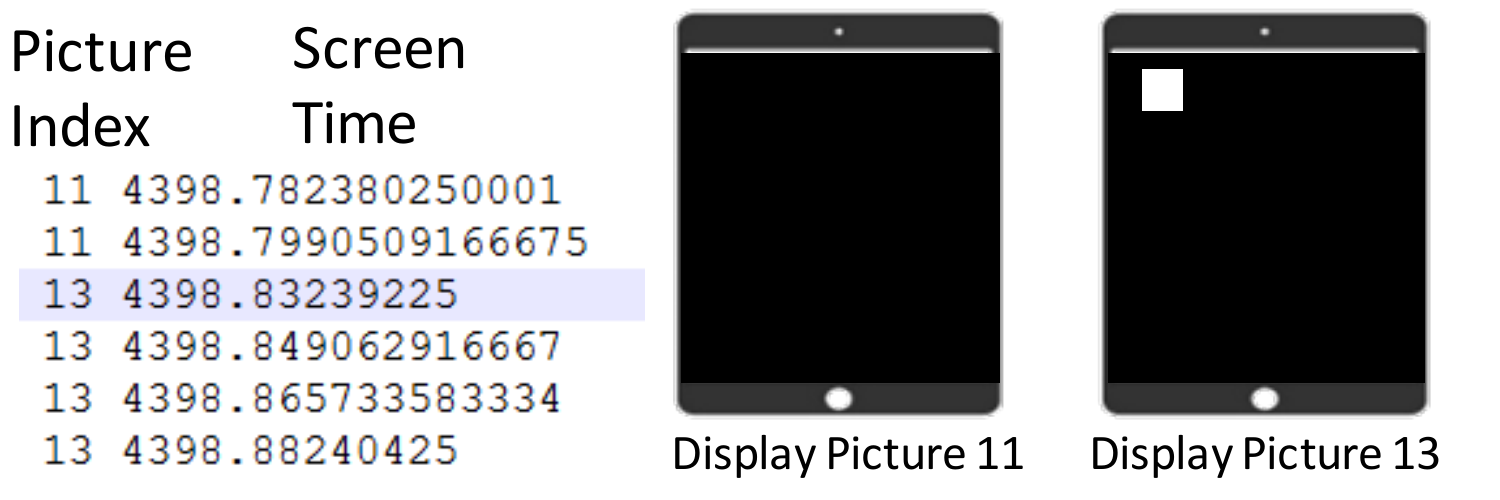}
     
      \caption{{Example for acquiring $s_i$, the time when the i-th stimulus presents on the screen. Picture 11 is a black image, and Picture 13 is the image with a left stimulus. The first stimulus shows up when Picture 13 is displayed. As a result, in this example, $s_1=4398.8324$ s. } 
      }
       
      \label{fig:screen}
  \end{figure}

To evaluate the synchronization error between the screen timestamps and the recording timestamps, we placed the device in front of a mirror and ran a 40-saccade task. With the mirror, we can identify the recording frame in which each of the 40 stimuli appears first. In Fig.~\ref{fig:recording}, for example, the first stimulus was presented in Frame 85. With the 40 frame indices and the associated recording timestamps $t_j$, we can translate these indices into time instants $r_i$ (ms), $i=1,\ldots,40$. In Fig.~\ref{fig:recording}, $r_1\approx 4398.8322$ s. Similarly, from the screen timestamps, we can obtain the time $s_i$ when the $i$-th stimulus is shown on the screen. Figure \ref{fig:screen}, shows $s_1$ to be approximately 4398.8324 s. 

If the timestamps were all accurate, the stimulus appearing on the screen would be captured by the next camera frame. In this case, $r_i-\frac{1000}{60}<s_i\leq r_i$, since the time difference between two frames is $\frac{1000}{60}$ ms in a 60-fps recording. If the errors in the recording timestamps and the screen timestamps are $D_r$ and $D_s$, respectively, the relationship becomes $r_i+D_r-\frac{1000}{60}<s_i+D_s\leq r_i+D_r$. That is, $r_i-\frac{1000}{60}<s_i-D\leq r_i$ where $D\defeq D_r-D_s$.

From the recording timestamps, we can only find one time instant $\tilde{r}_i(D)$ as a function of $D$ that satisfies $\tilde{r}_i(D)-\frac{1000}{60}<s_i+D\leq \tilde{r}_i(D)$. In other words, if each recording timestamp is denoted as $t_j$ where $j$ denotes the frame index as in Fig. \ref{fig:recording}, then $\tilde{r}_i(D):=\min \{t_j | t_j\geq s_i+D\}$. If our estimated synchronization error $\hat{D}$ is correct, we will have $\tilde{r}_i(\hat{D})=r_i$. As a result, we can then estimate $D$ by finding 
\begin{equation}
\hat{D}=\argmin_D \sum_i |\tilde{r}_i(D)-r_i|.
\label{eq:drift}
\end{equation}
With careful app design, we can ensure $\sum_i |\tilde{r}_i(\hat{D})-r_i|=0$. That is, our estimated synchronization error is correct.

\begin{figure}[b]
     \vspace{-9pt}
      \centering
      \includegraphics[width=\columnwidth]{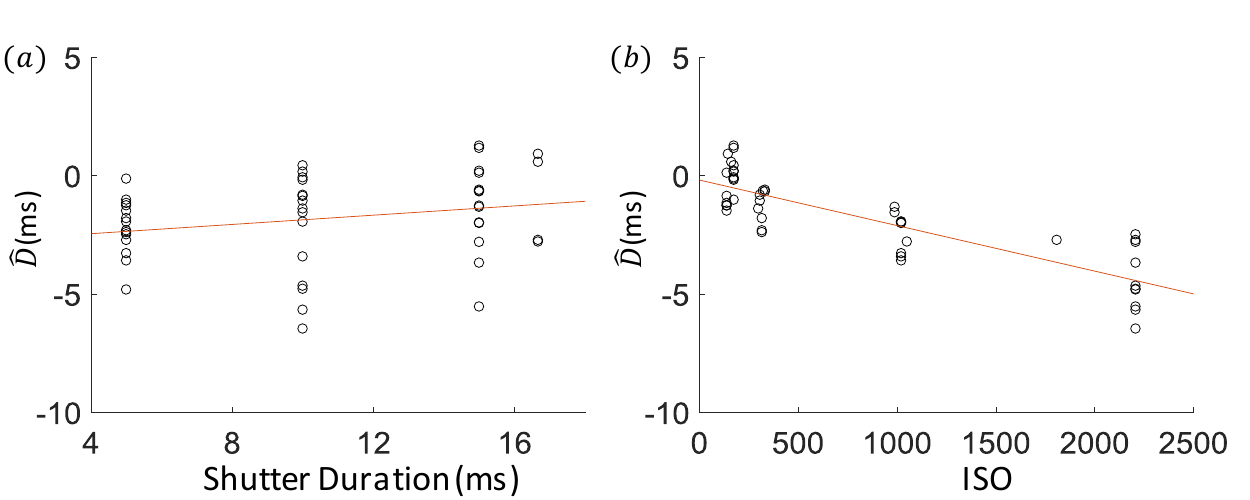}
     
      \caption{{The estimated synchronization error as a function of (a) shutter duration and (b) ISO. Each dot denotes one recording.} 
      }
      \vspace{-12pt}
       
      \label{fig:ISO}
  \end{figure}

We observed that an iOS camera changes its shutter duration and ISO based on the lighting condition, which may affect the accuracy of the recording timestamps. We showed in Fig.~\ref{fig:ISO} that the shutter duration does not affect the synchronization error while ISO is positively correlated with the absolute value of the synchronization error. As a result, we set the shutter duration to 16 ms, which is close to the maximum duration $1000/60$ ms in a 60-fps recording, to allow for adequate light. To bound the absoulte synchronization error to be within 5 ms, we restrict the ISO values to be less than 1000 by asking the subject to move to a brighter environment if the automatically determined ISO exceeds 1000. 

\end{document}